# On the Friedel sum rule in *ab initio* calculations of optical properties

D.V. Knyazev[1,2,*], P.R. Levashov[1,2]

[1] Joint Institute for High Temperatures RAS, Izhorskaya 13 bldg 2, Moscow 125412, Russia
[2] Moscow Institute of Physics and Technology (State University), Institutsky lane 9, Dolgopudny city, Moscow Region 141700, Russia

We investigate the influence of technical parameters in dynamic electrical conductivity calculations by the Kubo-Greenwood formula on the value of the so-called sum rule. We propose a possible explanation of the slight overestimation of the sum rule in most of our results.

## 1. Introduction

The Friedel sum rule is often mentioned in the papers on the *ab initio* calculations of optical properties [1]. The sum rule is used to check the validity of the *ab initio* calculation. In this work we consider some technical details connected with the sum rule. This work is generally based on the results published in our previous paper [2].

Our work is organized as follows. At first we discuss the theoretical derivation of the sum rule (Section 1). Then we consider different technical implementations of the Kubo-Greenwood formula (Subsections 3.1 – 3.5). Then the influence of the implementation on the sum rule and on the dynamic electrical conductivity curves is examined in Subsections 3.6 and 3.7 respectively. The possible explanation of the discrepancy between the calculated sum rule and unity is present in Subsection 3.8.

## 2. Theoretical derivation of the sum rule.

First of all, it should be noticed, that the sum rule is the exact relation. The sum rule may be derived from the very foundations of the quantum mechanics ([3]). The only condition implied is the completeness of the basis. In our case we have for $\hat{A} = \hat{x}$:

$$\frac{\hbar^2}{m}\langle i|i\rangle = 2\sum_j (\varepsilon_j - \varepsilon_i)|\langle i|\hat{x}|j\rangle|^2 = \frac{2\hbar^4}{m^2}\sum_j \frac{|\langle i|\nabla|j\rangle|^2}{(\varepsilon_j - \varepsilon_i)}.$$

This is also called the Thomas-Reiche-Kuhn sum rule [4]. If we now multiply this relation by the Fermi-weight $f_i$ (occupancies in the case of thermal equilibrium) and sum over all $i$ values we will obtain:

$$N = \frac{2\hbar^2}{m}\sum_{j>i}(f_i - f_j)\frac{|\langle i|\nabla|j\rangle|^2}{(\varepsilon_j - \varepsilon_i)}.$$

Here $N$ is the total number of occupied states. We can come to the continuous (over frequency) case by introducing simultaneously $\delta$-function and frequency integration:

$$N = \frac{2\hbar^3}{m}\int_0^\infty \sum_{i,j}(f_i - f_j)\frac{|\langle i|\nabla|j\rangle|^2}{(\varepsilon_j - \varepsilon_i)}\delta(\varepsilon_j - \varepsilon_i - \hbar\omega)d\omega.$$

Neglecting inessential details (such as spin degeneracy and averaging over three spatial directions) this is the sum rule for the Kubo-Greenwood formula [5].

This sum rule should be valid for any potential (even if it is physically meaningless), for any number of atoms (even for 1 atom), for any ionic configuration (even if it is incorrect for the given temperature). The only condition implied is the completeness of the basis. The validity of the sum rule does not guarantee the correctness of the obtained conductivity values.

## 3. Implementations of the Kubo-Greenwood formula.

We have checked the sum rule for the curves from Fig. 2 in our paper [2]. The integration range was from 0.005 eV up to 10 eV. The broadening of the $\delta$-function was 0.1 eV both for 108 and 256 atoms curves. 15 ionic configurations were taken into account.
The sum rule for 256 atoms is 1.03101.
The sum rule for 108 atoms is 0.912193.

This seems to be strange, because the sum rule is the exact relation and it should be valid for any number of atoms. To figure out the details of such sum rule behavior we have performed an additional investigation. To obtain the results quickly the investigation was performed for one ionic configuration (step 299). The number of atoms was 256. Firstly we examined the influence of the different technical realizations of the Kubo-Greenwood formula on the result. The summary may be found in the Table 1.

The value of the sum rule (step 299, $\Delta E = 0.1$ eV) is 1.02871.
The value of the sum rule (step 299, $\Delta E = 0.01$ eV) is 1.04298.

## 3.1. "Divide by frequency" and "divide by energy difference" options.

The first thing that may spoil the sum rule is the broadening of the $\delta$-function. The broadening itself does not affect the sum rule, because the area "under the $\delta$-function" is the same as the area under the broadened $\delta$-function.

However, in the Kubo-Greenwood formula (we mean its version presented in the paper by Desjarlais [1] and in our paper [2]) the $\delta$-function is multiplied by the $\frac{1}{\hbar\omega}$ function. This will affect the sum rule because $\frac{1}{\hbar\omega}$ function is convex downwards, and the different parts of the broadened $\delta$-function are multiplied not by $\frac{1}{\hbar\omega_{ij}}$ but by different values which do not compensate each other (Fig. 1a). We may even predict that the area will be overestimated.

This problem may be eliminated if we replace the conventional Kubo-Greenwood expression:

$$\sigma_1(\omega) = \frac{2\pi e^2 \hbar^2}{3m^2 \omega \Omega} \sum_{i=1}^{N}\sum_{j=1}^{N}\sum_{\alpha=1}^{3} \left[F(\varepsilon_i) - F(\varepsilon_j)\right] \left|\langle \Psi_j | \nabla_\alpha | \Psi_i \rangle\right|^2 \delta(\varepsilon_j - \varepsilon_i - \hbar\omega)$$

by the other one:

$$\sigma_1(\omega) = \frac{2\pi e^2 \hbar^3}{3m^2 \Omega} \sum_{i=1}^{N}\sum_{j=1}^{N}\sum_{\alpha=1}^{3} \left[F(\varepsilon_i) - F(\varepsilon_j)\right] \frac{\left|\langle \Psi_j | \nabla_\alpha | \Psi_i \rangle\right|^2}{(\varepsilon_j - \varepsilon_i)} \delta(\varepsilon_j - \varepsilon_i - \hbar\omega).$$

Here we multiply each broadened $\delta$-function not by changing $\frac{1}{\hbar\omega}$ function, but by constant $\frac{1}{\hbar\omega_{ij}}$ value (Fig. 1b). This will prevent the sum rule from being influenced. Further we will mark the conventional formula by *divide by frequency*, and the new version by *divide by energy difference*. Obviously, in the limit of zero broadening both formulas should give the same result.

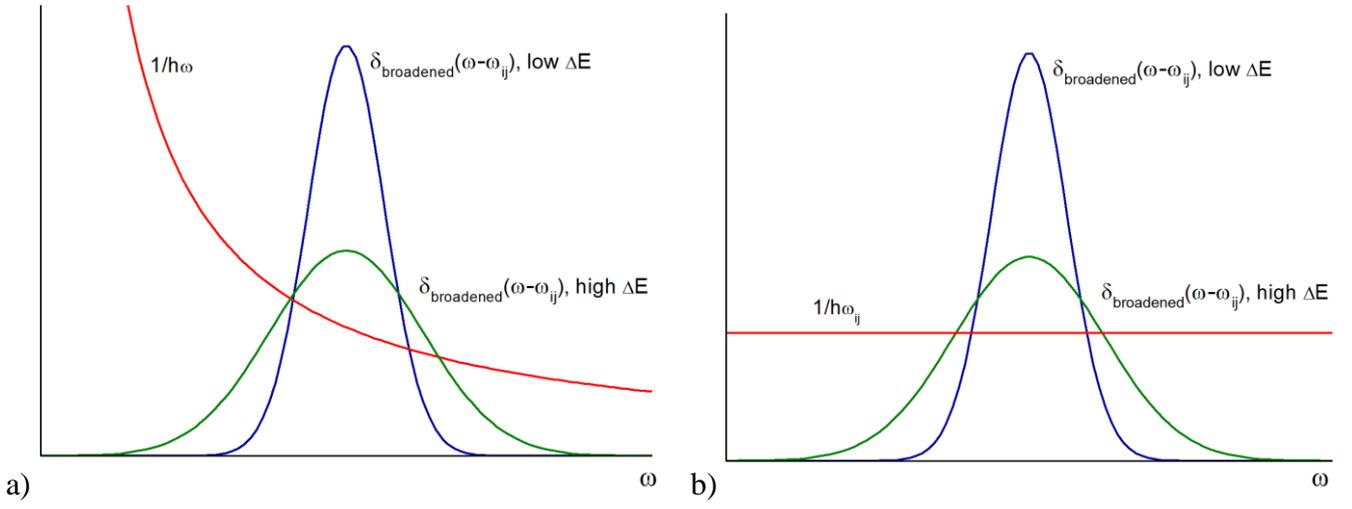

**Fig. 1.** The difference between "divide by frequency" and "divide by energy difference" options. a) "Divide by frequency" option, the broadened $\delta$-function is multiplied by convex downwards $\frac{1}{\hbar\omega}$ function, the area under the curve depends on the broadening. b) "Divide by energy difference" option, the broadened $\delta$-function is multiplied by the constant $\frac{1}{\hbar\omega_{ij}}$ value, the area under the curve does not depend on the broadening. The dimension of the vertical axes is inessential.

This affects the sum rule as follows (Table 1):

The sum rule (step 299, $\Delta E = 0.1$ eV, "divide by frequency") gives 0.976066.
The sum rule (step 299, $\Delta E = 0.01$ eV, "divide by frequency") gives 1.0357.

### 3.2. Zero correction option.

Another thing that may affect the sum rule is the zero frequency limit. If we take the conventional Kubo-Greenwood formula:

$$\sigma_1(\omega) = \frac{2\pi e^2 \hbar^2}{3m^2 \omega \Omega} \sum_{i=1}^{N}\sum_{j=1}^{N}\sum_{\alpha=1}^{3}\left[F(\varepsilon_i)-F(\varepsilon_j)\right]\left|\langle\Psi_j|\nabla_\alpha|\Psi_i\rangle\right|^2 \delta(\varepsilon_j-\varepsilon_i-\hbar\omega),$$

we may see that the broadened peaks close to the zero frequency "send the part of their area" to the negative frequencies. Moreover, some "negative peaks" make contribution to the conductivity: due to the broadening of the $\delta$-function some peaks with $\varepsilon_i > \varepsilon_j$ are also allowed, the difference $f(\varepsilon_i) - f(\varepsilon_j) < 0$ makes their contribution negative. This is illustrated by Fig. 2a.
The solution is to modify the Kubo-Greenwood formula as follows:

$$\sigma_1(\omega) = \frac{2\pi e^2 \hbar^2}{3m^2 \omega \Omega} \sum_{i=1}^{N}\sum_{j=1}^{N}\sum_{\alpha=1}^{3}\left|F(\varepsilon_i)-F(\varepsilon_j)\right|\left|\langle\Psi_j|\nabla_\alpha|\Psi_i\rangle\right|^2 \delta(\varepsilon_j-\varepsilon_i-\hbar\omega)$$

Here we have replaced $f(\varepsilon_i) - f(\varepsilon_j)$ (which may be both positive and negative) by $\left|f(\varepsilon_i) - f(\varepsilon_j)\right| > 0$. This returns missing area under the curve to the positive frequencies. This may also correspond to the following idea: we build the even $\sigma_1(\omega)$ function both for positive and negative frequencies and smooth it by the broadening of the $\delta$-function.
Further we will mark this new Kubo-Greenwood expression with *zero correction* label. Obviously, the differences between zero correction and the absence of zero correction are noticeable only at high enough broadenings $\Delta E$, and disappear in the limit of zero broadening. The use of the zero correction should increase the sum rule.

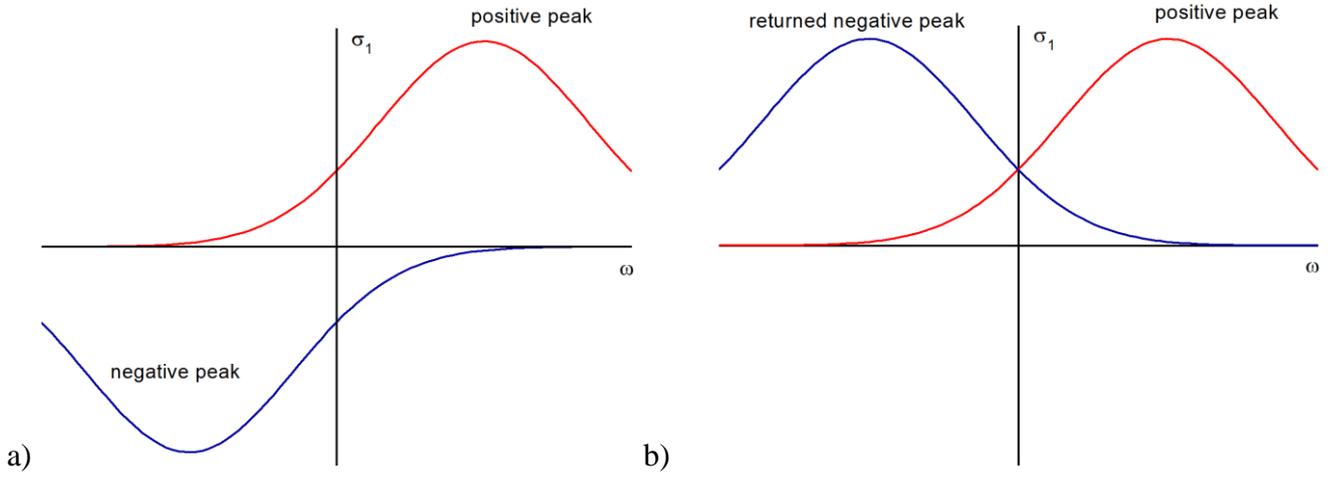

**Fig. 2.** The zero correction option illustration. a) No zero correction. Some area of the positive peak comes to the negative frequencies, negative peak gives its negative contribution to the area. b) Zero correction. Some area of the positive peak that came to negative frequencies is returned back by the returned negative peak.

The influence on the sum rule (Table 1):

The sum rule (step 299, $\Delta E = 0.1$ eV, "divide by frequency", zero correction) gives 1.25122 (huge value, see Fig. 4 and comments to it).
The sum rule (step 299, $\Delta E = 0.01$ eV, "divide by frequency", zero correction) gives 1.04918.

The influence of the zero correction combined with "divide by energy difference" (Table 1):
The sum rule (step 299, $\Delta E = 0.1$ eV, "divide by energy difference", zero correction) gives 1.0396
The sum rule (step 299, $\Delta E = 0.01$ eV, "divide by energy difference", zero correction) gives 1.0395
It may be easily seen that here the sum rule does not depend on the broadening.

### 3.3. Sum rule at high frequencies.

Only the finite range of frequencies (from 0.005 eV up to 10 eV) is included in the estimation of the sum rule. The missed contribution of the high frequency tail may be estimated as follows. At high frequencies $\sigma_1(\omega) \sim \dfrac{1}{\omega^2}$ (Fig. 3). So $\sigma_1(\omega > 10\,\text{eV}) \approx \dfrac{\sigma_1(10\,\text{eV}) \cdot (10\text{eV})^2}{\omega^2}$.

And the contribution of the tail may be estimated as:
$$S_{\text{tail}} = \dfrac{2m\Omega}{\pi e^2 N_e} \int_{10\,\text{eV}}^{+\infty} \sigma_1(\omega) d\omega \approx \dfrac{2m\Omega}{\pi e^2 N_e} \int_{10\,\text{eV}}^{+\infty} \dfrac{\sigma_1(10\,\text{eV}) \cdot (10\text{eV})^2}{\omega^2} d\omega = \dfrac{2m\Omega}{\pi e^2 N_e} \sigma_1(10\,\text{eV}) \cdot (10\text{eV}) \approx 0.05$$

In fact, this value is large enough. It only increases the value of the sum rule which is already larger than 1.

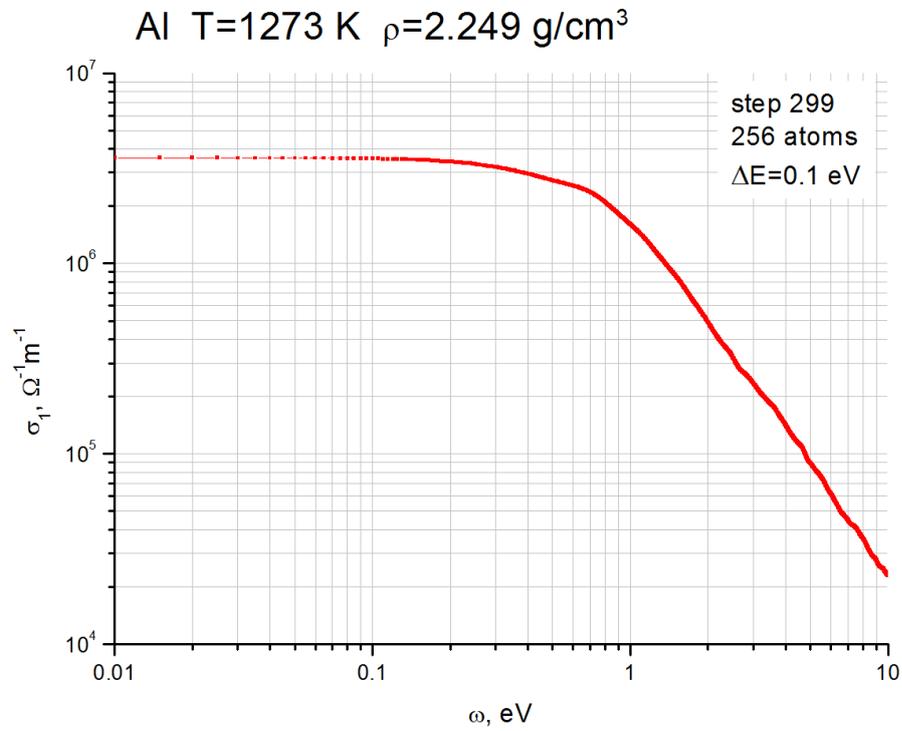

**Fig. 3.** The dependence of the dynamic electrical conductivity on the frequency. Here the Kubo-Greenwood formula is "divided by frequency" and no zero correction is included. The high frequency tail is close to $\sigma_1(\omega) \sim \dfrac{1}{\omega^2}$ asymptotics.

### 3.4. The influence of the integration step.

We may imagine that the integration step is not small enough, and the estimation of the integral is rather rough. It is not true. The frequency step in our paper [2] was 0.005 eV. We decreased it to 0.0005 eV. This did not influence the results greatly (less than 0.0001 changes of the sum rule).

### 3.5. The influence of the energy cut-off.

The calculation for the same ionic configuration with the energy cut-off 400 eV was also available. The sum rule was checked for it too (Table 1):

The value of the sum rule (step 299, $\Delta E = 0.1$ eV, ENCUT 400 eV) is 1.02964,
The value of the sum rule (step 299, $\Delta E = 0.01$ eV, ENCUT 400 eV) is 1.04425.

So the values do not change a lot in comparison with those obtained for ENCUT 200 eV.

### 3.6. The sum rule

The results on the sum rule are generalized in the following Table 1:

**Table 1.** The values of the sum rule for the different implementations of the Kubo-Greenwood formula. The values are given for the step 299. The number of atoms is 256.

|  | $\Delta E = 0.1$ eV | $\Delta E = 0.01$ eV |
|---|---|---|
| ENCUT 200; divide by frequency; no zero correction | 1.02871 | 1.04298 |
| ENCUT 200; divide by energy difference; no zero correction | 0.976066 | 1.0357 |
| ENCUT 200; divide by frequency; zero correction | 1.25122 | 1.04918 |
| ENCUT 200; divide by energy difference; zero correction | 1.0396 | 1.0395 |
| ENCUT 400; divide by frequency; no zero correction | 1.02964 | 1.04425 |

Here we come to the following conclusions: the values of the sum rule at low broadenings do not depend on the implementation of the Kubo-Greenwood formula (the differences are less then 0.02). The values at high broadenings may depend on the implementation of the Kubo-Greenwood formula. If we use "divide by energy difference" and zero correction options simultaneously, the calculated sum rule does not depend on the broadening. So we should use the values at the low broadenings to investigate, whether the sum rule is valid or not (in fact, this could be expected before this investigation). And we notice, that the sum rule value is systematically higher then 1 (in fact, the extra tail of 0.05 should be added).

### 3.7. The dynamic electrical conductivity.

How do the options "divide by energy difference" and zero correction influence the curves of dynamic electrical conductivity? This is shown in Fig. 4 for step 299. The number of atoms is 256.

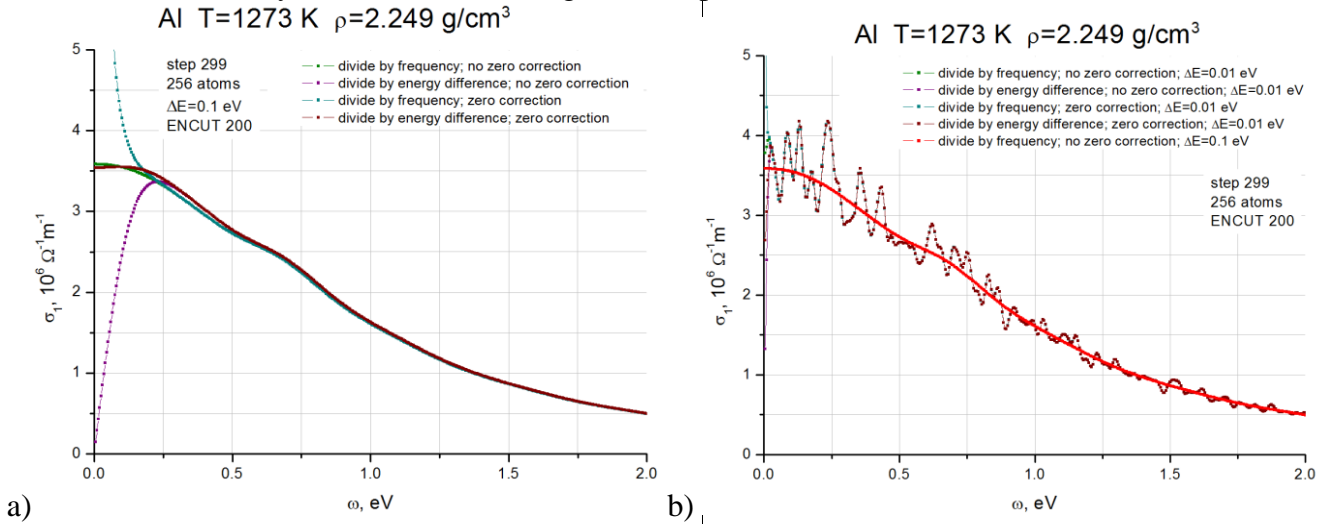

**Fig. 4.** The influence of different implementations of the Kubo-Greenwood formula on the dynamic electrical conductivity. a) The high broadening of the $\delta$-function. b) The low broadening of the $\delta$-function.

The results for the broadening 0.1 eV are shown in Fig. 4a. It may be seen that the use of only the "divide by energy difference" option or of only zero correction option leads to the bad behavior at low frequencies (abrupt drop and steep growth respectively). However the use of both options simultaneously yields good results, close to those obtained with the conventional Kubo-Greenwood formula implementation. It should be also noticed that the difference between different implementations is noticeable only at the frequencies lower then ~2-3 $\Delta E$.

At the low broadening $\Delta E = 0.01$ eV (Fig. 4b) all implementations give the same results. The curve with the high broadening for conventional implementation ("divide by frequency" and no zero correction) smoothes the oscillating curve well. The same may be done using the options "divide by energy" and zero correction simultaneously. The latter approach has additional advantage: here the sum rule does not depend on the broadening.

**3.8. The influence of the k-point number.**

Then what is the reason of the discrepancy between the obtained sum rule value and the unity? Different technical implementations of the Kubo-Greenwood formula were examined above and do not explain this discrepancy. We should also notice that for the pseudopotential used (ultrasoft pseudopotential, for details see our paper [2]) we have previously confirmed the ortonormality of the wave functions. We find the only explanation – the incompleteness of the basis. This may be improved by:
1) increment of the NBANDS parameter – it seems to be not true, because all the bands with significant occupations were taken into account, the sufficient number of empty states was added.
2) increment of the energy cut-off ENCUT, to represent each state better – we have checked this above.
3) increment of the **k**-point number. It seems to be the most reasonable solution. The use of the $\Gamma$-point only seems to result in an incomplete basis.

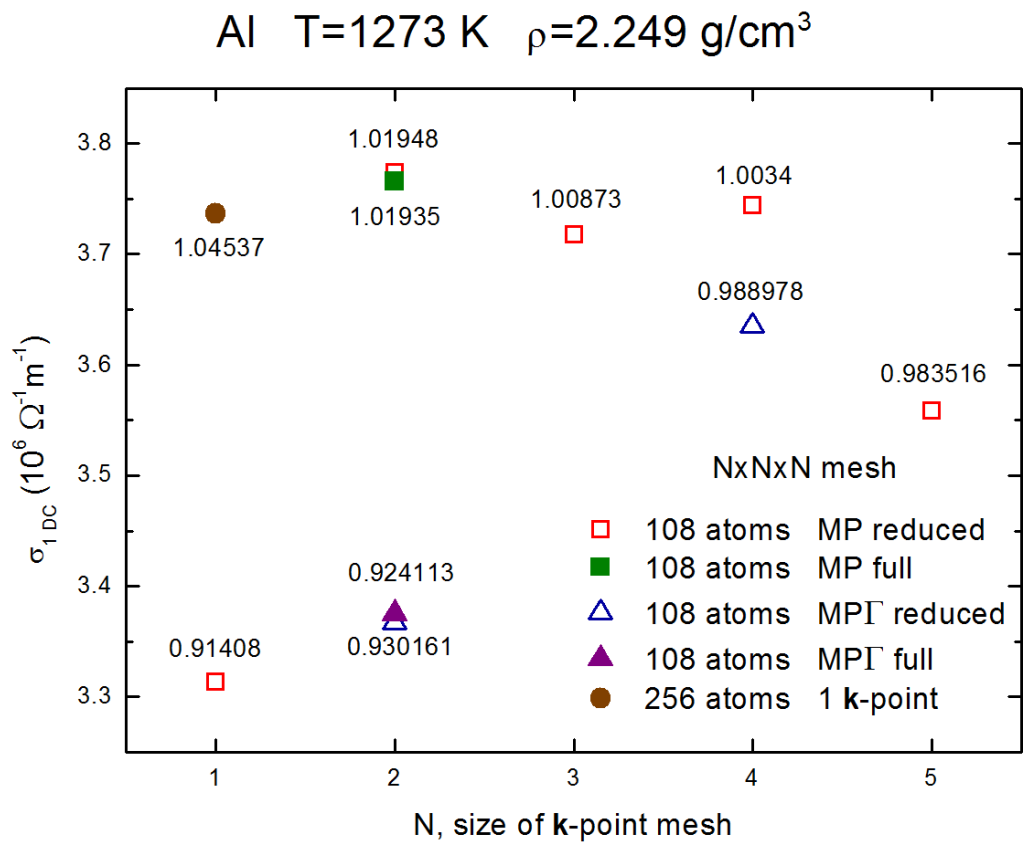

**Fig. 5.** The dependence of the results on the number of **k**-points. The calculated value of the sum rule is written near each point. The additional high frequency tail value of about 0.05 should be added to receive the full sum rule value.

To test this, we took Fig. 17 from our paper [2] (that with the investigation of influence of the **k**-point number on the result) and calculated the sum rule for each point. The results are shown in Fig. 5. The value of the sum rule is written near each **k**-point. The integration range was from 0.005 eV up to 10 eV. In fact, when looking at these values, one should remember about the additional high frequency tail of about 0.05. The number of ionic configurations is 15 for each point in the graph.

It may be seen (with some optimism) that the larger is the **k**-point number, the closer is the sum rule to unity. The values of the sum rule for $\Gamma$-point only both for 108 and 256 atoms do not equal 1 (the value for 256 seems to be even farther from unity than that for 108 atoms).

## 4. Conclusions

In conclusion we will summarize the main issues:
- the influence of the different implementations of the Kubo-Greenwood formula on the sum rule value was examined;
- for one particular implementation (with options "divide by energy" and zero correction) the sum rule value does not depend on the $\delta$-function broadening;
- even for a small broadening the sum rule differs from unity;
- the incompleteness of the basis seems to be the reason of this behavior;
- it seems that the problem may be improved by the increment of the **k**-point number.